\documentstyle[12pt]{article}

\advance\textwidth 2cm
\advance\oddsidemargin -1cm

\newcommand\beq{\begin{equation}}
\newcommand\eeq{\end{equation}}
\def\beqa{\begin{eqnarray}}
\def\eeqa{\end{eqnarray}}
\def\bega{\begin{array}}
\def\enda{\end{array}}
\def\be{\[}
\def\ee{\]}

\def\sp{\hspace{.8cm}}

\def\vb{\vphantom{b}}
\def\bq{{\bf q}}
\def\bp{{\bf p}}
\def\q#1#2{ { q^{#1\vb}_{#2} } }
\def\p#1#2{{p^{#1\vb}_{#2}}}

\def\sH#1{{\cal H}^{(#1)}}

\def\non{{\nonumber}}
\def\l{{\langle}}
\def\r{{\rangle}}
\def\c#1{_{,#1} }  

\def\dag#1{{\dagger_#1}}
\def\m#1{{\mu^{(#1)}}}
\def\phc#1{{\phi^{(#1)*}}}
\def\ph#1{{\phi^{(#1)}}}
\def\ps#1{{\psi^{(#1)}}}

\def\a{\alpha}
\def\b{\beta}

\def\la{\lambda}
\def\w{\omega}
\def\const{{\rm const}}

\def\arcsinh{{\rm arcsinh}}
\def\d{\partial}

\begin{document}

\title{\vspace{-1.5cm}
Unitary Equivalence of the Metric and Holonomy Formulations of
2+1 Dimensional Quantum Gravity on the Torus}
\author{Arlen Anderson\thanks{arley@physics.mcgill.ca}
\\Department of Physics\\ McGill University\\ Ernest
Rutherford Building\\
Montr\'eal PQ Canada H3A 2T8\\and\\
Blackett Laboratory\thanks{present address}\\
Imperial College\\Prince Consort Rd.\\
London SW7 2BZ England}
\date{October 12, 1992\\Revised December 31, 1992}
\maketitle

\vspace{-14cm}
\hfill McGill 92-19

\hfill Imperial/TP/92-93/02

\hfill gr-qc/9210007
\vspace{12cm}

\begin{abstract}
Recent work on canonical transformations in quantum mechanics is applied to
transform between the Moncrief metric formulation and the Witten-Carlip
holonomy formulation of 2+1-dimensional quantum gravity on the torus. A
non-polynomial factor ordering of the classical canonical transformation
between the metric and holonomy variables is constructed which preserves
their classical modular transformation properties. An extension of the
definition of a unitary transformation is briefly discussed and is used
to find the inner product in the holonomy variables which makes the canonical
transformation unitary. This defines the Hilbert space in the Witten-Carlip
formulation which is unitarily equivalent to the natural Hilbert space in
the Moncrief formulation. In addition, gravitational theta-states arising
from ``large'' diffeomorphisms are found in the theory.
\end{abstract}
\newpage

\section{Introduction}

The theory of 2+1-dimensional quantum gravity for the spatial topology of
the torus is quantum mechanical rather than field theoretic. As a
consequence, it can be exactly solved and serves as an excellent toy model
in which to study time in quantum gravity. Among the several treatments of
2+1-quantum gravity, there are essentially two contrasting classes of
formulation: that of Witten\cite{Wit,oth1} in terms of holonomies and that
of Moncrief\cite{Mon,oth2} in terms of the Arnowitt-Deser-Misner (ADM)
metric decomposition and the York extrinsic time. The most striking
distinction between them is that Witten's formulation is dynamically
sterile---the holonomies are constants of the motion---while, in Moncrief's
formulation, the 2-metric evolves as the York time progresses. This
difference highlights the ambiguity in defining time in quantum gravity.
The natural question is whether the two theories are equivalent, and, if
so, what this means about time.

Carlip\cite{Car1} has proven the classical equivalence of the two
formulations by defining variables which reflect the holonomy of the torus
and then giving the canonical transformation between his variables and the
metric variables of Moncrief. He finds that the canonical transformation is
a time-dependent one which trivializes Moncrief's Hamiltonian and thereby
transforms the time-dependent metric variables to time-independent ones.
Carlip goes on to discuss the quantum equivalence of the formulations by
undertaking the formidable task of factor-ordering the classical canonical
transformation. This procedure is not well-controlled; it is difficult to
find an ordering that takes one between two chosen Hamiltonians. Indeed,
Carlip does not obtain the Moncrief Hamiltonian, but rather one close to
it, which he calls a Dirac square-root\cite{Car2}.

The purpose of this paper is to use recent work on canonical
transformations in quantum mechanics\cite{And} to prove the unitary
equivalence of the metric and holonomy formulations. A general canonical
transformation can be implemented in quantum mechanics as a product of
elementary canonical transformations, each of which has a well-defined
quantum implementation. This is briefly reviewed in Section \ref{canon}. In
general the sequence of elementary canonical transformations that takes one
between two Hamiltonians classically is not the same sequence which does so
quantum mechanically because of terms that arise from factor ordering. This
implies that in general there is no simple (i.\ e.\ polynomial) factor
ordering of the classical general canonical transformation between two
Hamiltonians which relates their quantum variables. This is the reason that
attempting to factor order a classical result is uncontrolled: the
practical restriction to polynomial factor orderings means that one can
only relate those Hamiltonians for which the classical and quantum sequence
of elementary canonical transformations are essentially the same.

The quantum canonical transformation between the metric and holonomy
variables of 2+1-gravity on the torus will be found explicitly. First, in
Section~\ref{scl}, the classical solution of the metric super-Hamiltonian
and the classical transformation to Carlip's holonomy variables is
reviewed. In Section~\ref{sm2f}, a sequence of elementary canonical
transformations which transform the metric super-Hamiltonian to a massive
relativistic free particle is given. The sequence of transformations which
complete the trivialization of the super-Hamiltonian is then given in
Section~\ref{sf2t}, followed in Section~\ref{st2h} by the sequence which
transforms the variables of the trivialized super-Hamiltonian into those of
Carlip. A non-polynomial factor ordering of the classical canonical
transformation is found. Carlip\cite{Car1,Car2} emphasizes the importance of
the modular
transformation properties of the metric and holonomy variables. In
Section~\ref{smod}, the quantum canonical transformation between the metric
and holonomy variables is shown to preserve their classical modular
transformation properties.

In Section~\ref{unitary}, the familiar definition of a unitary
transformation as a linear norm-preserving isomorphism of a Hilbert space
onto itself is extended to transformations between Hilbert spaces. Given
a canonical transformation and the measure density for the inner product of
the original Hilbert space, a transformed measure density defining the
inner product of the new Hilbert space is found which preserves the values
of inner products between states. This is used to construct the measure
density of the inner product in the holonomy variables from the natural
measure density in the metric variables. This defines the Hilbert space in
the Witten-Carlip holonomy formulation. With this choice
of Hilbert space, the metric and holonomy formulations of 2+1-dimensional
quantum gravity on the torus are unitarily equivalent.

In Section~\ref{theta}, the role of modular transformations is considered
from the standpoint of the wavefunction. It is found that, in analogy to
the $\theta$-vacua of Yang-Mills, there are $\theta$-states in 2+1-gravity
on the torus which arise from ``large'' diffeomorphisms. Finally,
in the conclusion, the implications about the
nature of time following from the quantum equivalence of the metric and
holonomy formulations of 2+1-gravity on the torus are discussed.
In
an Appendix, solutions of the metric super-Hamiltonian constraint
are constructed using the canonical transformations found in
Section~\ref{sm2f}.

\section{Canonical Transformations}
\label{canon}

The use of canonical transformations in quantum mechanics has been
recently system\-atized\cite{And}.  This allows quantum systems to be
solved by transforming to simpler systems whose solutions are known.
Since time-dependent canonical transformations are of interest, it
is useful to work with the super-Hamiltonian in extended phase space, where
$(\q{}0,\p{}0)$ have been adjoined to the spatial variables.
A canonical transformation $P$ is an operator transformation between two
super-Hamiltonians
$$P\sH{0}P^{-1}={\cal H}^\prime.$$
This gives the solutions of $\sH{0}\ps{0}=0$ in terms of those of ${\cal
H}^\prime \psi^\prime=0$ as $\ps{0}=P^{-1}\psi^\prime$. It has been
conjectured that all integrable super-Hamiltonians $\sH{0}$ can be reduced
to trivialized form, where ${\cal H}^\prime=p_0$. There may be more than
one independent reduction to triviality, and together these give all
independent solutions of $\sH{0}$.

Canonical transformations exist independent of the choice of Hilbert space.
All solutions of one super-Hamiltonian, including non-normalizable ones,
are transformed to solutions of the other super-Hamiltonian. The conditions
under which a canonical transformation is a unitary transformation are
discussed in Section~\ref{unitary}. Canonical transformations are useful
even when they are not unitary\cite{And2}. In such cases, it is most
natural to impose boundary conditions and normalizability at the level of
$\sH{0}$, rather than upon the solutions of ${\cal H}^\prime$.

For completeness, a brief review of the use of canonical transformations
in quantum mechanics follows.
In principle, a general quantum canonical
transformation can be implemented as a product of three elementary
canonical transformations\cite{And}:  similarity (gauge) transformations, point
canonical transformations, and the interchange of coordinates and momenta.
Each of these is characterized by its transformation of the phase space
variables and its action on the wavefunction.  Since the transformations
are canonical, the transformed variables may simply be substituted in the
Hamiltonian for the original variables, respecting the original operator
ordering.

A similarity transformation is a shift of the momentum by a function of
the coordinate
\beq
p=p^a -if(q^a)_{,q^a}, \sp q=q^a ,
\eeq
where the comma indicates differentiation with respect to the subscript
and latin alphabet superscripts indicate the generation of the
transformation.
Under this transformation, the wavefunction changes by a factor
\beq
\ps{a}(q)=e^{-f(q)}\ps{0}(q).
\eeq
This transformation is often referred to as a
gauge transformation because of its role in the theory of a
particle interacting with an electromagnetic field.

A point canonical transformation is a change of variables
\beq
q=f(q^a), \sp p={1\over f(q^a)_{,q^a}}p^a.
\eeq
This transformation cannot always be expressed in terms of a single
exponential of a function, so it is denoted symbolically by its action on
the coordinate as $P_{f(q)}$.  Its action on the wavefunction is
\beq
\ps{a}(q)=P_{f(q)}\ps{0}(q)=\ps{0}(f(q)).
\eeq
Each occurence of $q$ is simply replaced by $f(q)$.

The interchange of coordinate and momentum is
\beq
p=-q^a, \sp q=p^a .
\eeq
This is implemented by the Fourier transform operator
\beq
I={1\over (2\pi)^{1/2} } \int_{-\infty}^\infty dq e^{iqq^a},
\eeq
and the wavefunction is transformed as
\beq
\ps{a}(q)=I\ps{0}(q)={1\over (2\pi)^{1/2} } \int_{-\infty}^\infty dq'
 e^{iq'q}\ps{0}(q').
\eeq

Using the interchange operator, analogs of the similarity and point
canonical transformations which involve functions of the momentum are
obtained by conjugation.  The composite similarity transformation is
\beq
q=q^a +if(p^a)_{,p^a}, \sp p=p^a ,
\eeq
and it acts on the wavefunction
\beq
\ps{a}(q)=e^{-f(p)}\ps{0}(q)=Ie^{-f(q)}I^{-1}\ps{0}(q).
\eeq
The composite point canonical transformation is
\beq
p=f(p^a), \sp q={1\over f(p^a)_{,p^a}}q^a
\eeq
and is given by $P_{f(p)}=I P_{f(q)} I^{-1}$.

Many-variable transformations can be constructed by using
many-variable functions in the elementary canonical transformations
above.  Since independent variables commute, all the variables not being
acted on may be treated as constant parameters.  For example, the point
canonical transformation $\exp(i\a q_1 p_1)$ generates the conformal
scaling
\beq
p_1=e^{-\a}p_1^a, \sp q_1= e^{\a} q_1^a,
\eeq
where
\be
\ps{a}(q_1,q_2)=\exp(i\a q_1 p_1)\ps0(q_1,q_2)=\ps0(e^\a q_1,q_2).
\ee
If $\a=\ln q_2$, then
\beq
p_1={\p{a}1 \over \q{a}2}, \sp q_1 =\q{a}2 \q{a}1.
\eeq
When this transformation acts on $(\q{}2,\p{}2)$, it is a similarity
transformation
\beq
\p{}2 = \p{a}2 +{\q{a}1\p{a}1\over \q{a}2},\sp \q{}2= \q{a}2.
\eeq
The wavefunction is transformed
\beq
\ps{a}(q_1,q_2)=\exp(i(\ln q_2) q_1 p_1)\ps0(q_1,q_2)=\ps0(q_1 q_2,q_2).
\eeq

\section{Classical Solution}
\label{scl}

To understand the relationship between the quantum and classical canonical
transformations between the metric and holonomy formulations of 2+1-gravity
on the torus, it is important to have both the quantum and classical
solutions at hand. The classical evolution has been discussed in detail
from a different perspective by Hosoya and Nakao\cite{HoN}.

The super-Hamiltonian in the metric variables\cite{oth2} is the
analog of the Wheeler-DeWitt equation for 2+1-gravity on the torus,
\beq
\label{sH0}
\sH0=-(\q{}0 \p{}0)^2 +\q{2}2(\p{2}1 +\p{2}2)=0.
\eeq
The ADM Hamiltonian
\beq
\label{MH}
H={\sqrt{\q{2}2(\p{2}1 +\p{2}2)} \over \q{}0 }
\eeq
found by Moncrief\cite{Mon} is a square-root of this super-Hamiltonian in a
sense discussed by Carlip\cite{Car2}.

The classical solution of the metric super-Hamiltonian follows from
the equations of motion
\beq
\bega{cc}
\dot q_0 = -2{\q{2}0} \p{}0 & \dot p_0= 2\q{}0 {\p{2}0}  \\
\dot q_1 = 2{\q{2}2} \p{}1 & \dot p_1= 0 \\
\dot q_2 = 2{\q{2}2} \p{}2 &\sp \dot p_2= -2\q{}2 ({\p{2}1}+{\p{2}2}),
\enda
\eeq
where the dot signifies differentiation with respect to the affine parameter
$t$.
Clearly, $\p{}1$ is a constant.
The general solution for the other variables is found to be
\beqa
\label{clsol}
\q{}0 &=& e^{-2 c_0 (t-t_0)}, \nonumber \\
\p{}0 &=& c_0 e^{2 c_0 (t-t_0)}, \nonumber \\
\q{}1 &=& {c_0  \over \p{}1 }\tanh( 2c_0 (t-t_0) +c_2) +c_1,  \\
\q{}2 &=& {c_0 \over \p{}1 \cosh( 2c_0 (t-t_0) +c_2) }, \nonumber \\
\p{}2 &=& -\p{}1  \sinh( 2c_0 (t-t_0) +c_2). \nonumber
\eeqa
There are only five constants of integration because in
addition to the six equations of motion there is one constraint, and this
fixes one constant of integration.
(Note that the degenerate solutions of the equations of motion, for which
the momenta vanish and the
coordinates are constant, are included in this solution as the special
case $\p{}1\rightarrow 0$, $c_0\rightarrow 0$, $c_0/\p{}1\rightarrow
\const$.)
By eliminating $t$ in favor of $\q{}0$, these
become
\beqa
\label{clsol2}
\p{}0 &=& {c_0 \over \q{}0}, \nonumber \\
\q{}1 &=& {c_0  \over p_1 }{1-\b^2  \q20 \over 1+\b^2  \q20}
  +c_1,  \\
\q{}2 &=& {2\b c_0 \q{}0 \over \p{}1 (1+\b^2  \q20) }, \nonumber \\
\p{}2 &=& -{\p{}1 (1-\b^2 \q20) \over 2\b  \q{}0}, \nonumber
\eeqa
where $\b=e^{-c_2}$.  Comparing with Carlip\cite{Car1},
the correspondence between the integration constants and Carlip's
holonomy variables is
\beqa
\label{ic}
p_1 &=& -2 a \la \nonumber \\
c_0 &=& -(a \mu -\la b)  \\
c_1 &=& {1\over 2} \left( {\mu\over \la} +{b\over a} \right), \nonumber \\
c_2 &=& \ln {\la\over a}. \nonumber
\eeqa

Carlip defines $(a,\mu)$ and $(b,-\la)$ to be pairs of canonically
conjugate variables.  Using the Poisson bracket defined in terms
of the holonomy variables,
\[ \{f,g\}_h = \d_a f \d_\mu g -\d_\mu f \d_a g -\d_b f \d_\la g +\d_\la f
\d_b g,  \]
the brackets among the integration constants in (\ref{ic}) can be
calculated.  It is found that $\{c_1,\p{}1 \}_h=2$, $\{ c_2, c_0\}_h =2$
and all other brackets vanish.  This implies that the transformation is
an extended canonical transformation\cite{LaL}. That is, the
transformation is canonical, but the momenta have been rescaled during
the transformation, so the brackets between conjugate variables
are not equal to one.

\section{Metric Super-Hamiltonian to Free Particle}
\label{sm2f}

There is a straightforward sequence of elementary transformations which
reduce the metric super-Hamiltonian (\ref{sH0}) to a massive relativistic
free particle (in 1+1-dimensional Minkowski space). This allows the
solutions of $\sH{0}$ to be constructed from the plane-wave solutions of
the relativistic free particle. Of course, the constraint $\sH{0}\ps{0}=0$
can be solved directly by separation of variables, but it is instructive to
construct the wavefunction by canonical transformation. This will be done
in the Appendix.

Here, the sequence of
transformations and the super-Hamiltonian after each transformation
will be given. For
convenience, only the variable(s) changed by a transformation will be
stated. A superscript indicating the generation of the transformation will
change for all variables.

Beginning with the metric super-Hamiltonian, the time-dependence is
simplified by a point canonical transformation
\be
\p{}0 = e^{-\q{a}0} \p{a}0, \sp \q{}0 = e^{\q{a}0},
\ee
and a linear term in $\q{a}2\p{a}2$ is introduced by shifting the momentum
\be
\p{}2 = \p{a}2 -{i\over 2\q{a}2}
\ee
giving
\beq
\sH{a}=- {\p{a}0}^2 + {\q{a}2}^2{\p{a}1}^2 +{\q{a}2}^2 {\p{a}2}^2
-i\q{a}2 \p{a}2 +  {1\over 4}.
\eeq
The wavefunction is
\beq
\ps{a}=\q{-1/2}2 P_{e^{\q{}0}}\ps{0}.
\eeq

A conformal canonical
transformation is made to absorb the momentum $\p{a}1$ into the coordinate
$\q{a}2$
\be
\q{a}1 = \q{b}1 -{\q{b}2\p{b}2 \over \p{b}1}, \sp
\p{a}2= \p{b}1 \p{b}2 , \sp
\q{a}2 = {1\over \p{b}1} \q{b}2.
\ee
This gives the super-Hamiltonian
\beq
\label{sHb}
\sH{b}=-{\p{b}0}^2 + {\q{b}2}^2 + {\q{b}2}^2 {\p{b}2}^2 -i\q{b}2 \p{b}2
  +  {1\over 4}.
\eeq
The wavefunction is transformed
\beq
\ps{b}=e^{-i\ln(\p{}1)\, \q{}2\p{}2}\ps{a}.
\eeq

As a differential operator, the spatial part of (\ref{sHb}) is the operator
in the modified Bessel equation. It can be transformed to the operator in a
Gegenbauer equation by making the interchange
\be
\p{b}2 = -\q{c}2, \sp \q{b}2 = \p{c}2 .
\ee
The super-Hamiltonian becomes
\beq
\sH{c}=-{\p{c}0}^2 + {\p{c}2}^2 (1+ {\q{c}2}^2) +i\p{c}2 \q{c}2
+{1\over 4}.
\eeq
The new wavefunction is given by the Fourier transform
\beq
\ps{c}=I_2\ps{b}={1\over (2\pi)^{1/2}} \int_{-\infty}^{\infty}
d\q{b}2 e^{i\q{b}2 \q{c}2} \ps{b},
\eeq
where the subscript on $I_2$ indicates which variable the transform acts
on.

The super-Hamiltonian has not been reordered with the momentum operators
on the right to facilitate the application of the
transformation
\be
\q{c}2 = \q{d}2 +{i\over \p{d}2}.
\ee
As a quantum operator, this can be expressed as
\beq
\q{c}2 ={1\over \p{d}2}\q{d}2 \p{d}2.
\eeq
This is an especially useful composite similarity transformation.
It is the realization of a first-order
intertwining operator as a canonical transformation\cite{And},
and it is an example of a transformation whose
behavior is
simpler in the quantum than the classical context.
The transformed super-Hamiltonian is
\beq
\sH{d}=-{\p{d}0}^2 + (1+ {\q{d}2}^2){\p{d}2}^2  -i\q{d}2 \p{d}2
+{1\over 4}.
\eeq
The wavefunction is
\beq
\ps{d}=\p{-1}2 \ps{c},
\eeq
where $\p{-1}2$ is the integral operator inverse to $\p{}2=-i\d_{\q{}2}$.

The point canonical transformation
\be
\p{d}2 = {1\over \cosh \q{e}2} \p{e}2, \sp
\q{d}2 = \sinh \q{e}2
\ee
reduces the super-Hamiltonian to the massive relativistic free particle
in Minkowski space
\beq
\label{free}
\sH{e}=-{\p{e}0}^2 + {\p{e}2}^2   +{1\over 4}.
\eeq
The new wavefunction is
\beq
\ps{e} =P_{\sinh{\q{}2}} \ps{d}.
\eeq

The original wavefunction is found by inverting the sequence of canonical
transformations.  It is given by
\beq
\label{wf0}
\ps{0}= P_{\ln\q{}0}(\q{}2)^{1/2}e^{i\ln(\p{}1)\, \q{}2\p{}2}I_2^{-1}\p{}2
P_{\arcsinh{\q{}2}}\ps{e}.
\eeq
This will be evaluated in the Appendix.

The accumulated canonical transformation from the original
super-Hamiltonian to the relativistic free particle is
\beqa
\label{M2f}
\q{}0 &=& \exp( \q{e}0 )  \nonumber \\
\p{}0 &=& \exp( -\q{e}0 ) \p{e}0 \nonumber \\
\q{}1 &=& {\tanh( \q{e}2) \over \p{e}1} \p{e}2 +\q{e}1  \\
\p{}1 &=& \p{e}1 \nonumber \\
\q{}2 &=& {1\over \p{e}1 \cosh(\q{e}2)}\p{e}2 \nonumber \\
\p{}2 &=& -\p{e}1 \sinh(\q{e}2) - {3 i \over 2\q{}2}. \nonumber
\eeqa
Note that a non-classical term involving $\q{}2$ appears in $\p{}2$.
This will be discussed below.

\section{Trivializing the Free Particle}
\label{sf2t}

The next step in the transformation to the holonomy
variables is to trivialize the relativistic free particle super-Hamiltonian.
This can be done in many ways, but the one which leads to the most
classical transformation from the metric variables to triviality is
accomplished by the sequence
\beqa
\p{e}0=(\p{f}0)^{1/2}, &\sp& \q{e}0=2(\p{f}0)^{1/2} \q{f}0, \nonumber \\
\p{f}0=\p{g}0+{\p{g}2}^2 +{1\over 4}, &\sp& \q{f}2=\q{g}2 -2 \p{g}2
\q{g}0, \\
\p{g}0= -\p{h}0, &\sp& \q{g}0= -\q{h}0. \nonumber
\eeqa
(In Section~\ref{smod}, this will also be found to be the transformation which
preserves the classical modular transformation properties of the
variables.)
These make the full transformation
\beqa
\label{f2t}
\p{e}0 &=& (-\p{h}0 +{\p{h}2}^2+{1\over 4})^{1/2},\nonumber \\
\q{e}0 &=& -2 (-\p{h}0 +{\p{h}2}^2+{1\over 4})^{1/2} \q{h}0, \\
\p{e}2 &=& \p{h}2, \nonumber \\
\q{e}2 &=&\q{h}2 +2\p{h}2 \q{h}0. \nonumber
\eeqa
The resulting super-Hamiltonian is
\beq
\label{sHh}
\sH{h}=\p{h}0.
\eeq

As an aside, note that after the constraint $\p{h}0=0$ is applied,
Eq.~\ref{f2t} implies that
\beq
\label{p0e}
\p{e}0=\sqrt{{\p{e}2}^2 +1/4}.
\eeq
This signifies that the canonical transformation used here is producing
the solutions to this part of the full constraint.  This corresponds
classically to the ADM Hamiltonian (\ref{MH}).
A second reduction would produce the solutions to $\p{e}0=-\sqrt{{\p{e}2}^2
+1/4}$.  That more than one reduction is needed to obtain all solutions
is evident because the second order operator (\ref{free}) has been reduced
to one of first order (\ref{sHh}).

Substituting (\ref{f2t}) into (\ref{M2f}) gives
\beqa
\label{M2t}
\q{}0 &=& \exp( -2 (-\p{h}0 +{\p{h}2}^2+{1\over 4})^{1/2} \q{h}0 )  \nonumber
\\
\p{}0 &=& \exp(2 (-\p{h}0 +{\p{h}2}^2+{1\over 4})^{1/2} \q{h}0 )
(-\p{h}0 +{\p{h}2}^2+{1\over 4})^{1/2} \nonumber \\
\q{}1 &=& {\tanh( \q{h}2 +2\p{h}2 \q{h}0) \over \p{h}1} \p{h}2 +\q{h}1  \\
\p{}1 &=& \p{h}1 \nonumber \\
\q{}2 &=& {1\over \p{h}1 \cosh(\q{h}2 +2\p{h}2 \q{h}0)}\p{h}2 \nonumber \\
\p{}2 &=& -\p{h}1 \sinh(\q{h}2 +2\p{h}2 \q{h}0) - {3 i \over 2\q{}2}. \nonumber
\eeqa
There is a clear correspondence with (\ref{clsol}) with the identifications
\beqa
\label{ic2}
\q{h}0 &=& t-t_0, \nonumber \\
\p{h}1 &=& \p{}1, \nonumber \\
\p{h}2 &=& c_0,  \\
\q{h}1 &=& c_1, \nonumber \\
\q{h}2 &=& c_2. \nonumber
\eeqa

The most evident differences between (\ref{clsol}) and (\ref{M2t}) are a
non-classical term in $\p{}2$ and the dependence of $\q{}0$ and $\p{}0$ on
$\p{h}2$ ($=c_0$). Consider the latter first. When the super-Hamiltonian
constraint $\p{h}0=0$ is imposed on physical states, the argument of the
exponential in the formula for $\q{}0$ becomes $-2({\p{h}2}^2+{1\over
4})^{1/2} \q{h}0$, instead of the $-2\p{h}2 \q{h}0$ expected from
(\ref{clsol}) after the
identification $\p{h}2=c_0$. The difference lies in the quantum shift of
$1/4$ which arose because the metric super-Hamiltonian is equivalent to a
massive rather than a massless free particle. The effect of this shift is
to make time flow differently in the quantum theory than in the classical.

The $1/4$ shift also contributes $(2i\q{}2)^{-1}$ to the non-classical term
in $\p{}2$. When this contribution is removed, the remaining
$(i\q{}2)^{-1}$ is the quantity needed to adjust the momentum $\p{}2$ so
that it becomes self-adjoint in the Petersson metric on moduli space. This
is argued in Sections \ref{theta} and \ref{obs} to be the natural inner
product for 2+1-gravity in a torus universe. Arguably, the self-adjoint
momentum operator, being observable, is the quantity which should be
compared with the classical momentum. Thus, one sees that the $1/4$ shift
introduces a quantum modification to the physically observable momentum.

Further effects on the flow of time in the quantum theory arise because of
the non-commutativity of $\q{h}2$ and $\p{h}2$. If one wants to eliminate
the affine parameter $\q{h}0$ in favor of the physical time $\q{}0$, it is
necessary to factor the exponentials in the hyperbolic functions defining
the spatial coordinates and momentum. The Baker-Campbell-Hausdorff (BCH)
formula may be used to find
\beq
\label{BCH}
\exp(\q{h}2 +2\p{h}2 \q{h}0)=\exp(\q{h}2)\exp(-i\q{h}0)\exp(2\p{h}2
\q{h}0).
\eeq
An additional factor in the affine parameter has arisen.  Eliminating
the affine parameter in favor of $\q{}0$ gives
\beq
\label{BCH2}
\exp(\q{h}2 +2\p{h}2 \q{h}0) =\exp(\q{h}2)
\q{ (i/2 -\p{h}2)({\p{h}2}^2+{1\over4})^{-1/2} }0
\eeq
(Note that in this formula $\p{h}0$ has been set to zero. This implies that
the formula is valid only when applied to physical states which satisfy the
super-Hamiltonian constraint. This restriction is important because the
presence of $\p{h}0$ would prevent the simple device of raising $\q{}0$ to
a power to replace $\q{h}0$.)

Eq.\ \ref{BCH2} can be used to obtain a quantum analog of (\ref{clsol2}),
but the result gives no additional insight. The point is that the modified
relation between $\q{}0$ and $\q{h}0$, together with the BCH factor, has
significantly complicated the quantum analog of (\ref{clsol2}), giving the
dependence of the spatial variables on $\q{}0$. It is the non-polynomial
nature of these complications that stand in the way of factor-ordering the
classical result, as Carlip attempted.

There is an important conclusion about factor-ordering to be inferred here.
It deserves emphasis and will benefit from restatement: When one attempts
to factor order a classical solution of a problem, one implicitly assumes
that the time-dependence of the quantum version is essentially unchanged
from the classical. This is naively justifiable because time is a c-number
and commutes with the quantum position and momentum operators one is
ordering. Closer inspection of time-dependent canonical transformations in
quantum theory reveals however that details of the super-Hamiltonian
sensitively affect the relation between the affine parameter and the
physical time. For the time-dependence of the ordered quantum operator to
be the same as it is classically, the quantum relation between the affine
parameter and the physical time must be the classical relation. This occurs
only for special super-Hamiltonians.

There are of course additional obstacles to factor ordering involving
non-polynomial orderings of the spatial variables. Generally, both of these
problems arise when the sequence of canonical transformations trivializing
a super-Hamiltonian are different classically and quantum mechanically. As
a rule, one cannot rely on the naive approach of looking for a polynomial
factor-ordering of a classical formula to find a quantum version.

\section{Transformation to Holonomy Variables}
\label{st2h}

To continue the transformation to holonomy variables, it is necessary first
to rescale the momenta. It was observed classically that the transformation
between the metric and holonomy variables is an extended canonical
transformation for which the Poisson brackets among variables were not
preserved but multiplied by a factor of 2. Rescaling of the momenta is
considered a trivial canonical transformation by Landau and
Lifshitz\cite{LaL}, but it deserves a brief discussion because the quantum
implementation differs from the conventional classical treatment.

Classically, when the momenta are rescaled by a constant factor $P=kp, Q=q$,
the Hamiltonian is rescaled by the same factor\cite{LaL}
\beq
k(pdq -H dt) = P dQ -H' dt.
\eeq
The result is that
Hamilton's equations do not change, and the transformation is canonical.
An alternative procedure is to leave the Hamiltonian unchanged but to
rescale the Poisson bracket
\beq
\{q,p\}=1, \sp \{Q,P\}=k.
\eeq
Clearly, this has the same effect of preserving the equations of motion.
Quantum mechanically, however, only the second procedure is consistent. The
canonical commutation relations induce a relationship between the original
coordinate and momentum operators which must be respected. Since the
coordinate doesn't change, $p=-i\d_q=-i\d_Q$ and therefore $P=kp=-ki\d_Q$.
As well, the form of the differential operator for the super-Hamiltonian is
unchanged when passing from $q$ to $Q$, so wavefunctions are preserved
under the momentum rescaling
\beq
\psi(q)=\psi^\prime(Q).
\eeq

In the transformation to the holonomy variables, before making the
rescaling, an interchange is used on $\p{h}2$
\be
\p{h}2=-\q{i}2, \sp \q{h}2= \p{i}2.
\ee
This enables the rescaling transformation to act on the coordinate
$\q{h}2$.  The rescaling transformation
\be
\p{i}1 =2 \p{j}1, \sp \p{i}2 =2 \p{j}2,
\ee
with the coordinates unchanged leaves the super-Hamiltonian unchanged
\beq
\sH{j}=\p{j}0.
\eeq
Note the commutation relations for the new variables are now $[\q{j}{\a},
\p{j}{\a}]=i/2$ ($\a=1,2$).  An inverse interchange
\be
\p{j}2=\q{k}2, \sp \q{j}2= -\p{k}2
\ee
switches the coordinate back to a momentum.

The next series of transformations serve to rearrange the variables.   They
establish the correspondence with Carlip's holonomy variables.  There
is a large amount of freedom in choosing the canonical variables that are
associated to a given super-Hamiltonian.  The classical correspondence
(\ref{ic})
between the integration constants and the holonomy variables guides the
choice.
The transformations are
\beqa
  \p{k}1 = -\q{l}1, &\sp& \q{k}1 = \p{l}1, \nonumber \\
  \p{l}1= e^{-\q{m}1}\p{m}1, &\sp& \q{l}1 = e^{\q{m}1}, \nonumber \\
  \p{m}1= \p{n}1+{1\over 2}\p{n}2, &\sp& \q{m}2 = \q{n}2 -{1\over 2}
\q{n}1, \nonumber \\
  \q{n}1= \q{o}1 +\q{o}2, &\sp& \p{n}2 = \p{o}2 -\p{o}1  \nonumber \\
  \q{o}1=\ln \q{p}1, \quad \p{o}1 =\q{p}1 \p{p}1, &\sp&
  \q{o}2=\ln \q{p}2, \quad \p{o}2 =\q{p}2 \p{p}2, \nonumber \\
  \q{p}2=-\p{q}2, &\sp& \p{p}2 =\q{q}2 \nonumber
\eeqa
The accumulated transformations from the $h$-variables are
\beqa
\p{h}1 &=& 2\q{q}1\p{q}2, \nonumber \\
\q{h}1 &=& {1\over 2} \left({\q{q}2\over \q{q}1} - {\p{q}1\over \p{q}2}
\right),  \\
\p{h}2 &=& -(\q{q}1 \p{q}1 + \p{q}2 \q{q}2), \nonumber \\
\q{h}2 &=& \ln ({-\p{q}2 \over \q{q}1} ). \nonumber
\eeqa
Using (\ref{ic2}), this agrees with the classical correspondence (\ref{ic})
between the
integration constants and Carlip's holonomy variables, where
\beq
\q{q}1=a,\quad \p{q}1=\mu,\quad \q{q}2=b,\quad \p{q}2=-\la.
\eeq

Using these in (\ref{M2t}) and eliminating the affine parameter in favor
of the physical time with (\ref{BCH2}) gives the full transformation
from the metric
variables to the holonomy variables.
For notational convenience, let
$$\a=(\p{h}2-{i\over 2}) ({\p{h}2}^2 +{1\over 4})^{-1/2} , $$
$$\b=(\p{h}2- i) ({\p{h}2}^2 +{1\over 4})^{-1/2}, $$
One finds
\beqa
\p{}1 &=& 2\q{q}1 \p{q}2 , \non \\
\q{}1 &=& {1\over {\q{q}1}^2 +{\p{q}2}^2 \q{-2\b}0 } (\q{q}1 \q{q}2 -
  \p{q}1 \p{q}2 \q{-2\b}0), \\
\p{}2 &=& -\q{\a}0 ({\q{q}1}^2 - {\p{q}2}^2 \q{-2\b}0) -  {3i \over 2\q{}2},
  \non \\
\q{}2 &=& {1\over {\q{q}1}^2 +{\p{q}2}^2 \q{-2\b}0 } \q{-\a}0 (\q{q}1
   \p{q}1 +\p{q}2 \q{q}2) . \non
\eeqa
In the classical limit, with the $1/4$ shift term dropped,
both $\a$ and $\b$ go to one, so the transformation
agrees with that found by Carlip \cite{Car1}.  Quantum mechanically, $\a$
and $\b$ are
non-polynomial ordering terms modifying the time dependence as discussed above.
Also note that, classically,
if one drops the quantum shift $1/4$ from the constraint
(\ref{p0e}), one finds
\beq
\p{}0={-(a\mu-\la b)\over \q{}0}.
\eeq
This is the classical expression that Carlip found for the ADM-Moncrief
Hamiltonian (\ref{MH}) in terms of the holonomy variables.

\section{Modular Transformations}
\label{smod}

The metric super-Hamiltonian is invariant under the symmetry of
modular transformations
of the upper-half plane.  This symmetry has its origin in the
diffeomorphism invariance of the original 2+1-gravity theory, as will be
discussed in the next section.  Defining $q=\q{}1+i\q{}2$ and
$p=\p{}1 +i\p{}2$, classically the modular
transformations on the metric variables are generated by
\beqa
\label{mtm}
T_m:q \rightarrow q+1,&\sp& T_m:p\rightarrow p, \\
S_m:q \rightarrow -{1\over q},&\sp& S_m:p \rightarrow q^{\dagger^2}p .\nonumber
\eeqa
These correspond to the transformations on the classical holonomy $q$-variables
\beqa
\label{mth}
T_h:(\q{q}1,\p{q}1) \rightarrow (\q{q}1,\p{q}1-\p{q}2), &\sp& T_h:
(\q{q}2,\p{q}2) \rightarrow (\q{q}1+\q{q}2,\p{q}2), \\
S_h:(\q{q}1,\p{q}1) \rightarrow (\q{q}2,\p{q}2), &\sp& S_h:
(\q{q}2,\p{q}2) \rightarrow (-\q{q}1,-\p{q}1). \nonumber
\eeqa

In the quantum theory, it is evident that the modular transformations of
$q$ induce the correct corresponding transformations of $p$, in the
coordinate representation where $\p{}1=-i\d_\q{}1$ and $\p{}2=-i\d_\q{}2$.
Thus, the classical
modular symmetry is consistently implemented quantum mechanically.

It is not obvious however that the quantum $q$-variables defined in terms
of the transformations above have their classical modular transformation
properties. Modular invariance of the quantum super-Hamiltonian is clearly
maintained, but since the holonomy super-Hamiltonian is trivialized, it
imposes no direct condition on the transformation properties of its
variables. Carlip's primary requirement\cite{Car1,Car2} in his approach to
factor ordering the classical solution was to find the quantum
transformation which preserves the classical modular transformation
properties of the holonomy variables. This, together with the practical
restriction to polynomial orderings, led him to conclude that the metric
super-Hamiltonian had to be modified.  Above, a non-polynomial ordering
was found which transforms from the unmodified metric super-Hamiltonian
to the holonomy version.  It remains to check that the modular
transformation properties have been preserved.

The modular transformations (\ref{mtm}) and (\ref{mth}) are themselves
canonical transformations.  By finding their explicit representation as a
product of elementary canonical transformations, the problem of comparing
them is greatly simplified.  The canonical transformations producing the
modular transformations in the holonomy
variables are not difficult to find because they are linear canonical
transformations
\beqa
T_h &=& e^{i2\q{q}1\p{q}2} \\
S_h &=& e^{i2 \q{q}2\p{q}1} e^{-i2 \q{q}1\p{q}2} e^{i 2\q{q}2\p{q}1}  .
\nonumber
\eeqa
In the metric variables, in complex form $q=\q{}1+i\q{}2$, $p=\p{}1+i\p{}2$,
the canonical
transformations are again linear.  There is a subtlety because the
momentum operator conjugate to $q$ is $p^\dagger=-2i\d_q$.  Recognizing
that $(q,p^\dagger)$ and $(q^\dagger,p)$ are two independent (commuting)
sets of variables, one finds
\beqa
T_m &=& e^{i\p{}1} \\
S_m &=& e^{-iq^{\dagger^2}p/2}e^{-ip/2}e^{-iq^{\dagger^2}p/2}
e^{-iq^{2}p^\dagger/2}e^{-ip^\dagger/2}e^{-iq^{2}p^\dagger/2} \nonumber\\
&=& e^{-i[(\q{2}1-\q{2}2)\p{}1+2\q{}1\q{}2\p{}2]}e^{-i\p{}1}
e^{-i[(\q{2}1-\q{2}2)\p{}1+2\q{}1\q{}2\p{}2]}. \nonumber
\eeqa

Comparing these, the modular transformations are the same if
\beq
\p{}1=2\q{q}1\p{q}2
\eeq
and
\beq
(\q{2}1-\q{2}2)\p{}1+2\q{}1\q{}2\p{}2 =-2 \q{q}2\p{q}1.
\eeq
Both conditions are satisfied by the transformation from the metric to
holonomy variables.  The first is obvious as it is one of the
transformations.  The second follows from a computation most easily done by
first passing from the metric variables to the
$h$-variables and then to the $q$-variables.  One finds
\beqa
(\q{2}1-\q{2}2)\p{}1+2\q{}1\q{}2\p{}2 &=& \q{h}1 \p{h}1 \q{h}1 -{{\p{h}2}^2
\over \p{h}1} \\
&=&-2\q{q}2 \p{q}1. \nonumber
\eeqa

Thus, the quantum canonical transformation given here from the metric to
the holonomy variables preserves the modular transformation properties of
each.  A word of caution should be raised:  it is not difficult to find
other canonical transformations to other variables which have the correct
classical
correspondence and which preserve modular invariant quantities, but do
not have the desired modular transformation properties.  This shows that
there is important information contained in the transformation properties
of non-invariant quantities.

\section{Unitary Equivalence}
\label{unitary}

Having constructed the canonical transformation between the metric and
holonomy variables which preserves their classical modular transformation
properties, it is necessary to determine if the transformation is unitary.
If it is, this will complete the proof of the unitary equivalence of the
two quantum theories.

The familiar definition of a unitary transformation is a linear
norm-preserving isomorphism of one Hilbert space onto itself. This
definition can be naturally extended to a linear norm-preserving isomorphism
from one Hilbert space to another\cite{And2}. Canonical transformations are
not in themselves unitary as they are defined independent of the Hilbert
space structure and transform all solutions, not just normalizable ones, of
one super-Hamiltonian constraint to solutions of another. For a canonical
transformation to define a unitary equivalence, when restricted to act on
the Hilbert space of states of one theory, it must be a unitary
transformation to the Hilbert space of states of the other theory.

In proving the unitary equivalence of the metric and holonomy formulations
of 2+1-gravity on the torus, there is a difficulty because Witten (and his
successors) did not derive the inner product which defines the Hilbert
space of states in the holonomy variables. The holonomy-variable inner
product for which the two formulations are unitarily equivalent will be
constructed by requiring that the value of transition amplitudes computed
in the metric-variable inner product be preserved through the canonical
transformation. After checking that the kernel of the canonical
transformation does not lie in the Hilbert spaces, it is concluded that the
canonical transformation is a norm-preserving isomorphism of the Hilbert
spaces. The two theories are then equivalent. If one were to choose a
different modular invariant inner product in the holonomy formulation, the
two theories would be unitarily inequivalent.

An inner product is characterized by its measure-density $\mu(\bq,\bp)$
[$\bq=(q,q_0)$, $\bp=(p,p_0)$, where $q$, $p$ stand for all of the spatial
variables],
\beq
\l \phi| \psi \r_\mu \equiv \int dq \phi(q)^* \mu(\bq,\bp) \psi(q),
\eeq
where $\bp$ acts to the right.  Note that in general the measure density may
be operator-valued and time-dependent.  The standard inner product with trivial
measure density is given by $\mu=1$.

In general, when one makes a canonical transformation, if the value of the
inner product
is to be preserved, the measure density will transform.
If one has the
canonical transformation between solutions,
\be
\ps0=C\ps{a},
\ee
then formally one has
\beqa
\l \ph0 | \ps0 \r_{\m0} &=& \l C\ph{a} |\m0 | C\ps{a} \r_1  \\
&=& \l \ph{a} | C^\dag1 \m0 C |\ps{a} \r_1 \nonumber \\
&=& \l \ph{a} | \ps{a} \r_{\m{a}}. \nonumber
\eeqa
The transformed measure density is
\beq
\label{mtran}
\m{a}(\bq,\bp)=C^\dag1 \m0(\bq,\bp)C.
\eeq
Here, $C^\dag1$ is the adjoint of $C$ in the trivial measure density.

For functions of $q$ and $p$, the adjoint in the trivial measure density
is the operator formed by complex
conjugation and integration by parts (all boundary terms are assumed to
vanish, though this must be checked in specific examples).  The
``adjoints''
of the interchange operator and the point canonical transformation can be
computed by direct manipulation of inner products in which they appear.
They are found to be
\beqa
P_{f(q)}^\dag1 &\equiv& f^{-1}(q)\c{q} P_{f^{-1}(q)} \\
I^\dag1 &\equiv& I^{-1}. \non
\eeqa
(The factor $f^{-1}(q)\c{q}$ in the point canonical transformation
adjoint arises from the transformation of the $dq$ in the
measure.  This is the one-dimensional form.)
For canonical transformations involving $p_0$, the adjoint cannot be
taken because the inner product does not involve an integration over
$dq_0$.  Point canonical transformations of the time simply redefine the
variable one uses for the time label and are implemented
directly in the measure density by changing its time-dependence
accordingly.

In common practice, one usually only considers unitary transformations
between
a Hilbert space and itself.  In these cases, the measure density does not
change,
and one finds from (\ref{mtran}) that $C^\dagger C=1$, where
$C^\dagger=\m0^{-1} C^\dag1 \m0$ is the adjoint in the measure density
$\m0$ of the Hilbert space.
It should be emphasized that canonical transformations which are not naively
unitary, such as multiplication by a real function of the coordinate,
become so when the measure density is appropriately transformed.

There is a natural inner product in the metric formulation of 2+1-gravity
on the torus given by the
measure density for the Petersson metric on the upper-half plane,
$\m0=q_2^{-2}$,
\beq
\label{mip}
\l \ph{0} | \ps{0} \r_{\m0}= \int {dq_1 dq_2\over q_2^2} \phc{0} \ps{0}.
\eeq
The canonical
transformation from the holonomy to metric variables is summarized by
\beq
\ps0=C\ps{q}.
\eeq
It is convenient for presentation to decompose $C$ into three transformations
\be
C=C_1 C_2 C_3,
\ee
where
\beqa
\ps0 &=& C_1 \ps{e} \non \\
\ps{e} &=& C_2 \ps{h} \\
\ps{h} &=& C_3 \ps{q}. \non
\eeqa
The canonical transformations are then
\beqa
\label{ctran}
C_1 &=& P_{\ln q_0} q_2^{1/2} e^{i(\ln p_1) q_2 p_2} I_2^{-1} p_2
P_{\arcsinh q_2} \non \\
C_2 &=& P_{p_0^2} e^{i(p_2^2 +1/4)q_0} e^{\pi q_0 p_0} \\
C_3 &=& I_2^{-1} R_{1/2} I_2 I_1^{-1} P_{\ln q_1} e^{i p_2 q_1/2}
e^{-ip_1 q_2} P_{e^{q_1}} P_{e^{q_2}} I_2. \non
\eeqa
Here, $R_{1/2}$ scales the momenta by a factor of one-half. Its ``adjoint''
would act to
double any momenta appearing in the measure density.

Using (\ref{mtran}),  one can compute the transformed measure
density one transformation at a time.  The final result is that
\beq
\m{q}=-(\q{q}1 \p{q}1 +\p{q}2  \q{q}2).
\eeq
The inner product in the holonomy variables which preserves the value of
inner products (\ref{mip}) in the metric variables is then
\beq
\label{hip}
\l \ph{q} | \ps{q} \r_{\m{q}}= -\int d\q{q}1 d\q{q}2 \phc{q}
(\q{q}1 \p{q}1 +\p{q}2  \q{q}2) \ps{q}.
\eeq

Before concluding that the metric variable theory with inner product
(\ref{mip}) is unitarily equivalent to the holonomy variable theory with
this inner product, one must confirm that no states in the Hilbert space
are in the kernel of the transformation $C$. All of the transformations in
$C$ are invertible except for $p_2$ in $C_1$ whose kernel is spanned by the
function $1$. Transforming this function to find its expression in the
holonomy variables, one finds a function which is not modular invariant. It
is therefore not a member of the Hilbert space. As this is the only
function which is annihilated by the transformation $C$, all modular
invariant functions in the holonomy variables are mapped to modular
invariant functions in the metric variables and vice versa. (The
preservation of modular invariance is guaranteed by the considerations of
Section~\ref{smod}.)

The canonical transformation $C$ is a (linear) norm-preserving isomorphism
of the Hilbert spaces defined by the inner products (\ref{mip}) and
(\ref{hip}). Therefore, one may conclude that the metric and holonomy
formulations of 2+1-gravity on the torus are unitarily equivalent. If one
chooses a different modular invariant inner product in the holonomy
variables, for example, a trivial measure $\m{q}=1$, then the metric and
holonomy formulations would not be unitarily equivalent.

This emphasizes the important point that a quantum theory is not complete
until the Hilbert space is specified. Note that once one allows
operator-valued measure densities, it is not clear what criteria one uses
to choose among those in which the Hamiltonian is self-adjoint and which
are invariant under the necessary symmetries. This introduces a new
ambiguity into quantization.

\section{Metric from Holonomy Wavefunctions?}

Having proven the unitary equivalence of the metric and holonomy
formulations of 2+1-gravity on the torus, one might hope to use the
holonomy wavefunctions to construct the metric wavefunctions by simply
applying the canonical transformation $C$. This would be a significant
achievement because the metric wavefunctions are the weight-zero Maass
forms and are of interest to number theorists\cite{Iwa}. In principle, this
procedure is straightforward. There are few subtleties in applying $C$, even
with its time-dependence---the time is simply a
parameter in the transformations.  To illustrate the use of a canonical
transformation, in the Appendix
the transformation $C_1$ is applied to plane wave solutions of the
relativistic free particle to obtain the corresponding
(non-modular invariant) solutions of the metric super-Hamiltonian.

Unfortunately, there is no ``free lunch.'' The trouble is that the
canonical transformation transforms all solutions to the holonomy
super-Hamiltonian, that is, all time-independent functions, to solutions of
the metric super-Hamiltonian. The solutions of interest however are only
the normalizable modular invariant functions. It is straightforward to
define what one means by a modular invariant function in the holonomy
variables using the transformations (\ref{mth}), but to the author's
knowledge, these functions are not known explicitly. Hence, they cannot be
transformed to give explicit representations of the Maass forms.
Furthermore, it is likely that, were they known, their expression would not
be in closed form, but in the form of infinite series. Evaluation of the
various Fourier transforms involved in $C$ would then result in a series
expansion. It is not obvious that the subtleties of Maass forms would be
more transparent in this form. Granted, it is an
improvement to be
able to work with modular invariant functions which are
not constrained to satisfy a differential equation, but further
investigation is required.

\section{Theta-states for 2+1-quantum gravity}
\label{theta}

In deriving the metric super-Hamiltonian, the configuration space arises
from gauge-fixing the diffeomorphism invariance of 2+1-gravity. The
reduction from the space of all metrics on the torus to moduli space is
made by observing that every metric on the torus is conformal to one of
constant zero-curvature.  In particular, the metric can be expressed in the
form $ds^2=e^\rho |dx +\tau dy|^2$, where $\tau=q_1+i q_2$ is a complex
parameter.
Restricting attention to the tori of
zero-curvature having the metric $ds^2=|dx+\tau dy|^2$, fixes the ``small''
diffeomorphisms, i.e. those that are
continuously deformable to the identity. The
zero-curvature tori are classified by their moduli $\tau=q_1+i q_2$,
and the Teichm\"uller
space of the torus is the Poincar\'e upper-half plane $H$. From this, it
follows that the wavefunction will be a function of the moduli, and it is
natural to expect that the inner product will be that on the upper-half
plane.

In addition to small diffeomorphisms, there are also ``large''
diffeomorphisms which are not fixed by restricting to zero-curvature tori.
These are diffeomorphisms which are not continuously deformable to the
identity and correspond to Dehn twists: the action of cutting open the
torus along a homotopically non-trivial loop and twisting the end before
gluing the manifold back together again.

Large diffeomorphisms are the analog of the large gauge transformations in
Yang-Mills that give rise to the theta-vacua\cite{Jac}. The possibility of
theta-states in gravity has been discussed in the past\cite{Ish,HaW}. Until
recently\cite{GiL}, no explicit examples were known. Gravitational
theta-states are present in 2+1-quantum gravity on the torus, and they are
constructed below. Their existence was overlooked in earlier
treatments\cite{Car1,Car2,HoN}.

Since two tori related by a large diffeomorphism are physically equivalent,
their moduli do not represent distinct configurations. The action of the
large diffeomorphisms on moduli space is given by the mapping class group,
$\Gamma=SL(2,Z)/Z_2$. The physically distinct moduli lie in a fundamental
domain $H/\Gamma$ which, under the action of the mapping class group,
tesselates the upper-half plane.

A solution $\ps{0}$ of the metric super-Hamiltonian constraint on
Teichm\"uller
space must be ``periodized'' to account for the physical equivalence of
moduli in different copies of the fundamental domain. This is accomplished
by the method of images in which the wavefunction $\Psi^{(0)}(q)$ at a
point $q=\q{}1+i\q{}2$ in the fundamental domain is found by summing the
solution $\ps{0}$ evaluated at every image of $q$ under the mapping class
group, each weighted by some factor $\chi_\a$,
\beq
\label{Psi}
\Psi^{(0)}(q)= \sum_{\a \in \Gamma} \chi_\a \ps{0}(\a q).
\eeq
Contrary to naive expectation\cite{Car1,HoN}, the wavefunction is not
required to be invariant under the action of the mapping class group, but
rather it must transform as a representation of the group. This determines
the possible weights $\chi_\a$.

The situation is analogous to that first considered by Laidlaw and
DeWitt\cite{LaD}. They were studying the propagator for a particle in a
multiply-connected configuration space, but a related argument works for
the wavefunction and applies to covering groups of non-topological origin.
In the sum over images, the fundamental domain is a particular coset
representative $H/\Gamma$ that has been arbitrarily chosen. If a second
coset representative were selected to be the fundamental domain by acting
with an element of the mapping class group, the new wavefunction must be
unitarily equivalent to the original wavefunction. Thus, if the
wavefunction $\Psi^{(0)}$ is to be a (modular weight zero) scalar, it can
change at most by a scalar phase
\beq
\label{trnsf}
\Psi^{(0)}(\b q)=e^{-i\phi(\b)} \Psi^{(0)}(q),
\eeq
and every image in the sum must be changed by the same phase
\[ \sum_{\a \in \Gamma} \chi_{\a}\ps{0}(\a\b q)=
\sum_{\a \in \Gamma} e^{-i\phi(\b)}\chi_{\a\b}\ps{0}(\a\b q).
\]
This is true (though possibly not in the most general way) if
$e^{i\phi(\b)}\chi_{\a}=\chi_{\a\b}$. Assuming without loss of generality
that the weight in the fundamental domain is one, $\chi_e=1$, one finds
$\chi_{\b}=e^{i\phi(\b)}$. Since the mapping class group is non-abelian
while phases commute, the phases must form a one-dimensional unitary
representation of the abelianization of the mapping class group.

The mapping class group is generated by the two fundamental modular
transformations
\[ S:q \rightarrow -q^{-1}, \]
\[ V: q \rightarrow -(q+1)^{-1}, \]
where $V=TS$, in terms of $T$ used above. The modular transformations
satisfy the relations: $S^2=1$, $V^3=1$. Every element of the mapping class
group can be represented as an element of the free product of $S$ and $V$.
By assigning a phase to $S$ and to $V$ that is consistent with their
relations, a phase is assigned to each element of the mapping class group.
This phase is a character of the abelianization of the mapping class group,
$Z_2\times Z_3$. The possible phases are $(1,e^{i\pi})$ for $S$ and
$(1,e^{2\pi i/3}, e^{4\pi i/3})$ for $V$. Thus, there are potentially five
non-trivial scalar theta-states consistent with modular transformations, in
addition to the modular invariant wavefunction. Two of these are not
physically distinct however as the states formed with the $V$-phase
$e^{4\pi i/3}$ are the complex conjugate of those formed with $e^{2\pi
i/3}$. There are three non-trivial scalar theta-states.

Note that from (\ref{trnsf}) the wavefunction $\Psi^{(0)}$ must vanish at
the fixed points of a modular transformation if the phase associated with
that transformation is not unity. This defines the boundary conditions that
are associated with each of the theta-states. These boundary conditions are
analogous to those that arise, say, at a reflecting wall where a
wavefunction must vanish because $\phi(x)=-\phi(-x)$.

Theta-states were found by dropping the requirement that the wavefunction
be modular invariant. Further generalizations are possible by weakening
other assumptions. If the wavefunction $\ps0$ is not a scalar, but has a
vector/spinor index, as the spinor wavefunction of Carlip's Dirac
square-root does, non-abelian weights are possible. If the weight is a
matrix of the same dimension as that of the vector/spinor, by contracting
it with the wavefunction's index, a vector/spinor of the same dimension is
obtained, and the above argument may be repeated. It is found that the
weights must be in a unitary matrix representation of the mapping class
group of dimension equal to the dimension of the vector/spinor. There are
more unitary representations in higher dimensions than in one, so there
will be more theta-states, and many will have the novel feature of being
non-abelian. This does not happen in Yang-Mills theory and is (so far)
unique to gravitational theta-states.

In principle, one can also use non-abelian weights in higher-dimensional
unitary representations of the mapping class group even when $\ps0$ is a
scalar (or is of different dimension than the weight). This possibility was
raised by Hartle and Witt\cite{HaW}. The result of doing so is that the
wavefunction $\Psi^{(0)}$ is no longer a scalar but carries a group label,
forming a higher-dimensional representation of the mapping class group. The
inner product must accordingly be adjusted to remain a modular invariant,
so that the group label does not appear in transition amplitudes. The
physical significance of such an extended wavefunction is not clear.

\section{Observables}
\label{obs}

Since the eigenvalues of a complete set of independent observables
characterize the quantum state, it is useful to consider the observables
for 2+1-gravity in a toroidal universe.
An observable is an operator
which commutes with all of the
constraints, up to a function of the constraints. This definition may be
unfamiliar because one is not used to quantum mechanics in the presence of
constraints\cite{And3}.
If constraints are present, they must be preserved. Suppose a
wavefunction $\Psi$ satisfies the constraints $C_i$, and
$[A,C_i]=B_i$, then
\be
[A,C_i]\Psi=-C_i A\Psi=B_i\Psi.
\ee
If $A\Psi$ is to continue to satisfy the constraints, $B_i$ must be a
function of the constraints.

Here, the observables must commute with the super-Hamiltonian and be
modular invariant. Because the super-Hamiltonian is modular invariant,
there are no additional constraints. Inspection reveals that $\p{}1$ and
$H^2=\q{2}2(\p{2}1 +\p{2}2)$ commute with the metric super-Hamiltonian.
Since the configuration space is two-dimensional, they are complete.
Unfortunately, $\p{}1$ is not modular invariant. This means that its
eigenvalues will not be quantum numbers of the ``periodized'' quantum
states (\ref{Psi}). They are nevertheless useful because they characterize
the quantum states $\ps0$, appearing in the sum over images, which are
constructed in the Appendix.
It is an open problem (to the author's knowledge) to find
a complete set of modular invariant observables.

Following the *-algebra approach\cite{Ash}, an inner product of a
constrained system can be constructed by requiring that each of the
observables be self-adjoint. Lacking a complete set of modular invariant
observables, a physical inner product for the periodized quantum states
$\Psi^{(0)}$ cannot be constructed directly. It is instructional however to
construct the inner product for $\ps0$, before imposing the symmetry of
modular invariance. Since the coordinates and momenta satisfy the ordinary
commutation rules in the coordinate representation, they have the usual
conjugation properties.
Assume
that the measure density is solely a function of the coordinates.
Then, for $\p{}1$ to be self-adjoint, the measure
density must be independent of $\q{}1$ while, for $H^2$ to self-adjoint,
the measure density must be $\q{-2}2$.
The inner product is thus determined to be
\beq
\int {1\over \q{2}2} d\q{}1d\q{}2\, \psi^* \phi.
\eeq
This is the standard inner product on the Poincar\'e upper-half plane with
the Petersson metric, as it must be since $H^2$ is the covariant Laplace
operator on this space. Note that the measure is modular invariant, even
though this is not a priori obvious from the construction. This implies
that this is in fact the physical inner product. This construction may seem
superfluous since it was already known that the configuration space is the
moduli space of the torus and thus has this as its natural inner product.
In more general problems, however, an understanding of the configuration
space may not precede the construction of observables.

\section{Conclusion}

A sequence of elementary canonical transformations has been found that
trivializes the metric super-Hamiltonian of 2+1 dimensional quantum gravity
on the torus. A further sequence establishes the quantum canonical
transformation between the metric variables and the holonomy variables of
Carlip. The full transformation is a non-polynomial factor ordering of the
classical canonical transformation which preserves the classical modular
transformation properties of the metric and holonomy variables. The
procedure used here enabled a systematic derivation of the transformation.

The definition of a unitary transformation was extended to apply to
transformations between Hilbert spaces having inner products with different
(operator-valued) measure densities. Requiring that the canonical
transformation be such a unitary transformation from the natural Hilbert
space in the metric formulation to the Hilbert space in the holonomy
formulation gave a construction of the inner product (\ref{hip}) in the
holonomy formulation. This proves that the Witten-Carlip holonomy
formulation of 2+1 dimensional quantum gravity on the torus
with this Hilbert space is unitarily equivalent to the Moncrief
metric formulation with its natural Hilbert space. If a different inner
product were used in the holonomy variables, the formulations would be
inequivalent.

The primary motivation for studying 2+1-gravity on the torus is to
understand the different natures of time in the metric and holonomy
formulations. There is no mystery here, but rather an important lesson.
Amongst all possible variables equivalent under canonical transformation,
Witten fortuitously chose a set that were time-independent while Moncrief
did not. Mathematically, both choices are equally valid and equivalent.
Physically, it becomes a question of what one can measure with experimental
apparatus. Set aside the inconvenient fact that strictly speaking
neither the metric nor the
holonomy variables are observable since they are not modular
invariant. If one had at hand a device which measures holonomy, then as one
went out into one's 2+1-dimensional world, there would appear to be no
dynamics. If, on the other hand, one were to measure the moduli of the
manifold, one would find that they change as the volume of the universe
grows. If one went out with a device which measures something else, one
would find yet different dynamics.

The lesson is first a familiar one from classical mechanics:  when
solving a problem, one is free to choose the variables which make it
convenient to solve; the physics does not depend on the choice of variables.
Second, it is a reminder that ``time'' is not a coordinate label, but a
perception that follows from physical observation.  While it does not
matter what variables one chooses to formulate a theory, it makes all the
difference what quantities one measures experimentally.  It is easy to
confuse measurable quantities with the variables that represent them most
conveniently, but those same quantities can be expressed in different
variables with no loss of information.  Time, as a quantity whose passage
is inferred indirectly from the change of other measured quantities, is
no different:  it is not the coordinate $\q{}0$ which appears in the
super-Hamiltonian.  The challenge is to understand
the connection between
measurable quantities and the physical passage of time so that this
connection
can be taken over into the theory and be preserved as a relation
amongst variables.  Properly executed, the relation will persist no matter
what the choice of variables.

\subsection*{Acknowledgements}  I would like to thank A. Ashtekar, S.
Carlip,  J. Louko and R. Myers for interesting discussions.  This work was
supported in part by
a grant from the Natural Sciences and Engineering Research Council of Canada
and les Fonds FCAR du Qu\'ebec.
\\

\section*{Appendix}

The formal solution of the metric super-Hamiltonian can be constructed by
applying the canonical transformation $C_1$ of (\ref{ctran}) to the
relativistic free particle wavefunction.  As we are interested in the
stationary states of the metric super-Hamiltonian, we begin
with the familiar positive-frequency plane-wave solutions of the
relativistic free particle,
multiplied by a plane wave in $\q{}1$,
\beq
\ps{e}_{k,n}(q_1,q_2,q_0)=\exp(ikq_2-i\w q_0)\exp(2\pi i nq_1)
\eeq
where $\w=(k^2+1/4)^{1/2}$.  Strictly, any
function of $\q{}1$ gives a solution of $\sH{e}$; the plane wave
is chosen so that
$\ps{e}$ is an eigenstate of the (partial) observable $\p{}1$ discussed in
Section~\ref{obs}.   The original wavefunction is given
\beq
\ps0(q_1,q_2,q_0)= P_{\ln q_0}\q{1/2}2  e^{i\ln p_1 q_2 p_2} I_2^{-1} p_2
P_{\arcsinh q_2} \ps{e}(q_1,q_2,q_0).
\eeq

Focusing on the $q_2$-dependent part (suppressing the subscript 2),
the transformation
\beq
I^{-1} p P_{\arcsinh q} e^{ikq} ={1\over (2\pi)^{1/2}}
\int_{-\infty}^\infty dq' e^{-iqq'} (-i\d_{q'}) \exp(ik \arcsinh q').
\eeq
is to be evaluated.  Doing the $\d_{q'}$ derivative and using $\arcsinh q'=
\ln (q'+(q^{\prime\,2}+1)^{1/2})$ gives
\beq
{1\over (2\pi)^{1/2}}
\int_{-\infty}^\infty dq' e^{-iqq'} k { (q'+(q^{\prime\,2}+1)^{1/2})^{ik}
\over (1+q^{\prime\,2})^{1/2} } .
\eeq
The change of variables $u=q'+(q^{\prime\,2}+1)^{1/2}$ gives an integral
which integrates to a modified Bessel function, up to a constant $N$,
\beq
{k\over (2\pi)^{1/2}}
\int_{0}^\infty du \exp(-iq(u-{1\over u})/2) u^{ik-1}=N K_{ik}(q).
\eeq

Applying $\exp(i\ln(p_1) q_2 p_2)$ to this scales $q_2$ by a factor of
$p_1$, giving
\beq
\exp(i\ln(p_1) q_2 p_2) K_{ik}(q_2) \exp(2\pi i nq_1) = K_{ik}(q_2 p_1)
\exp(2\pi i nq_1).
\eeq
Acting with the operator $\p{}1$, this becomes
\beq
K_{ik}(2\pi n q_2) e^{2\pi i n q_1}.
\eeq
Including the final transformations, the wavefunction is
\beq
\ps0_{n,k}=N \q{1/2}2 K_{ik}(2\pi n q_2) e^{2\pi i n q_1} \q{i\w}0.
\eeq

{}From Section \ref{theta}, the full periodized wavefunction is
\beq
\label{psum}
\Psi^{(0)}(q_1,q_2,q_0)= \sum_{\a \in \Gamma} \chi_\a \ps{0}_{n,k}(\a q_1,\a
q_2,q_0),
\eeq
where $\chi_\a$ is a unitary one-dimensional representation of $Z_2\times Z_3$.
Thus, the solutions of the metric super-Hamiltonian have been constructed
using canonical transformations. It should be noted that the condition of
normalizability
has not been applied, and one does not expect every periodized sum
(\ref{psum}) to correspond to a normalizable wavefunction.
The normalizable wavefunctions are however expected to be among those
obtained by the periodizing procedure.

\end{document}